\documentclass[10pt]{article}

\def\doctitle{Privacy Issues of the W3C Geolocation API}
\def\docauthor{Nick Doty, Deirdre K. Mulligan and Erik Wilde}
\def\repdate{February 2010}
\def\irep{UC Berkeley School of Information Report 2010-038}

\usepackage[utf8]{inputenc}
\usepackage{verbatim}

\usepackage{graphicx}
\usepackage{fancyheadings}
\usepackage{lastpage}
\usepackage[margin=1in,letterpaper,includeheadfoot]{geometry}
\usepackage[a4paper={false},letterpaper={true},
colorlinks={true},
linkcolor={blue},
citecolor={black},
urlcolor={blue},
pdfstartview={Fit},
pdfview={Fit},
bookmarks={true},
pdftitle={\doctitle},
pdfsubject={\irep},
pdfauthor={\docauthor}]{hyperref}

\def\uri#1{{\tt \url{#1}}}
\def\code#1{{\tt #1}}

\begin{document}

\title{\doctitle}
\author{\docauthor\\
\href{http://www.ischool.berkeley.edu/}{School of Information, UC Berkeley}}
\date{\irep\\
\repdate\\
\ \\
Available at \uri{http://escholarship.org/uc/item/0rp834wf}}
\maketitle

\thispagestyle{empty}

\begin{abstract}

The W3C's Geolocation API may rapidly standardize the transmission of location information on the Web, but, in dealing with such sensitive information, it also raises serious privacy concerns. We analyze the manner and extent to which the current W3C Geolocation API provides mechanisms to support privacy. We propose a privacy framework for the consideration of location information and use it to evaluate the W3C Geolocation API, both the specification and its use in the wild, and recommend some modifications to the API as a result of our analysis.

\end{abstract}

\vfill
\tableofcontents
\newpage

\section{Introduction}\label{intro}

With the advent of mobile access to the Internet (early technologies, now mostly replaced by fully Internet-capable access, were i-mode and WAP), location-based services have become an increasingly popular part of the Web, from restaurant-finding to driving directions to social network location-sharing. The collection, use and disclosure of location information to provide such services implicates users' privacy and physical safety. Information about an individual's location routinely receives heightened privacy protection under law and in social interactions.

Currently the World Wide Web Consortium's Geolocation \emph{Application Program Interface (API)} that enables scripting code on a web page to access device location information is the leading candidate for standardizing the transmission of location information on the Web. The API can be rapidly deployed: Web developers can simply write scripting code to use this new browser functionality; no other part of web infrastructure need be updated. This is a distinct advantage over other possible mechanisms, such as extensions of the \emph{Hypertext Transfer Protocol (HTTP)}~\cite{dav03b}, the main protocol used for communications on the Web, or the development of a \emph{Uniform Resource Identifier (URI)} scheme for location~\cite{may09}.

Given the high likelihood of the broad adoption of the W3C Geolocation API and its expected influence on user privacy and security, the current ``last call'' working draft\footnote{The most up-to-date version is the editor's draft: \uri{http://dev.w3.org/geo/api/spec-source.html}} deserves detailed analysis.
The extent to which technical standards can and should provide the capacity for privacy and other values to be addressed is contested.
While few dispute the influence that technology and interface design have on privacy, the extent to which technology affirmatively should be designed to support privacy-protective outcomes is a live question.
The authors have a range of opinions on this topic and addressing this question as a general matter is beyond the scope of this paper.
However, we do believe that it is important to understand the strengths and limitations of the Geolocation API with respect to supporting privacy.
Without such an understanding it will be difficult to address location privacy on the Web whether through technology, law, norms or market forces.

In this report we analyze the manner and extent to which the current W3C Geolocation API provides mechanisms to support privacy. We first discuss prior standards efforts to address privacy concerns. Next, drawing on U.S. law and normative conventions around withholding and sharing location information, we outline the privacy and physical safety concerns raised by the wide availability of geolocation information. We then propose a privacy framework for the consideration of location information and use it to evaluate the W3C Geolocation API, both the specification and its use in the wild.  We also identify issues that fall outside the scope of the API but, given current market structures, raise additional privacy issues that should be addressed in a holistic consideration of location information privacy.  In conclusion, we suggest four modifications to the API that would provide users with greater support for their privacy and physical safety concerns.

\section{Standards and Privacy}\label{standards-and-privacy}

Attention to the effect of web standards on individual privacy began in earnest in late 1994 when heated debate arose at the Internet Engineering Task Force over proposed changes to the hypertext transfer protocol that standardized the ability of third-parties to set ``cookies'' in users browsers, but discouraged doing so~\cite{kri01}. A nascent network advertising trade objected to defaults that would impede their business model by limiting third-party ``cookies'' while privacy advocates, the press and regulatory bodies weighed in on the other side objecting to the invisibility of the data collectors, the lack of user control, and the risks of data aggregation~\cite{kri01}. Beginning in 1996 the role of web standards --- technical standards operating at the application layer that define, describe and support the operation of the World Wide Web --- in enabling privacy protection became an area of research, debate and standards activity with the formation of the World Wide Web Consortium's \emph{Platform for Privacy Preferences} Project (P3P)~\cite{cra02,cra97,kri01}. Shortly thereafter, the IETF's geographic location privacy (Geopriv) working group formed in recognition that ``the representation and transmission of [location] information has significant privacy and security implications.'' In response to these concerns, Geopriv has ``created a suite of protocols that allow such applications to represent and transmit such location objects and to allow users to express policies on how these representations are exposed and used.''\footnote{\uri{http://www.ietf.org/dyn/wg/charter/geopriv-charter.html}}

In a similar vein, efforts to consider the policy implications of rights expression languages analyzed their ability to support both the rights and limitations in copyright law~\cite{mul03b}. At a higher level, several techniques for identifying and addressing the impact of technical design on values have been offered with varying effect~\cite{fla08,mor03b,cla02}.
In the area of privacy specifically researchers have directed attention to the ways in which technical design alters norms of information flow~\cite{nis04}, removes structural barriers that provide de facto privacy protection~\cite{sur07}, increases visibility, transparency and exposure~\cite{coh08}, introduces persistent identifiers, facilitates monitoring and tracking and enables the collection and retention of information about individual users~\cite{mor03b}.

\section{Location Information Privacy}\label{location-privacy}

Information about location --- both real-time location as well as permanent locations (such as home address) --- is afforded special attention due to the consequences for both privacy and physical safety that may flow from its disclosure.

The heightened privacy and physical safety concerns generated by the collection, use and disclosure of location information are apparent in U.S. law that creates restrictive consent standards for its use and disclosure in the private sector in the context of telecommunications services (47 USC \S 222), court concern about the standards governing law enforcement access to real-time and historical location information from telecommunications providers under current law\footnote{\uri{http://www.eff.org/related/3494/pressrelease}}, as well as in limitations on the disclosure of department of motor vehicle records under federal law~\cite{dppa}, state laws allowing specific individuals to limit the disclosure of their address in public records~\cite{ale04} and the requirement to offer caller-id suppression services.

At the level of norms, the heightened sensitivity with which individuals regard location information is reflected in social practices as well as empirical data. Parents and educators routinely remind children not to give out their home address or their physical location. This advice carries through strongly in educational efforts aimed at digital youth, where a primary rule of the road is to never give out address or location information to strangers or post it publicly. It is also reflected in a trend to remove address information from phone directories, limit the availability of school contact lists, and other community efforts to protect the privacy of home addresses. Survey data confirms this evident sensitivity of location information privacy. For example, a recent representative survey of Californians found strong support for judicial intervention and due process before law enforcement are given access to historical location data (73\% supporting a law requiring ``police to convince a judge that a crime has been committed before obtaining location information from the cell phone company''~\cite{kin08b}). Qualitative work exploring the use of location-sharing platforms has found that a range of privacy and security concerns influence decisions to share.  For example, one study found that who was requesting information and the purpose for the request had greatest influence over decisions to disclose or withhold, with the user's location and activity being less important~\cite{con95}. Relatedly, location information has been identified as the most sensitive element of information shared within social networks~\cite{pat95}.

\section{Privacy Framework}\label{framework}

Acknowledging the privacy and safety considerations attending location information is helpful, however what is needed at this point is guidance about what issues can and should be addressed at the API level. Theoretical and prescriptive work on protecting privacy exists in several fields. The framework we develop below draws primarily from research in the law and computer and human interaction fields.

Much of the legal framework for protecting privacy, including location information privacy, derives from the Fair Information Practice Principles that are premised on the concept of ``informational self-determination'' and are substantiated in procedural rules that establish: individual rights of data subjects to control the collection and reuse of personal information, and access and correct or amend or delete personal information held by organizations; and, organizational obligations to minimize the collection of personal information, provide safeguards against misuse and unauthorized access and respect the rights of data subjects~\cite{oecd80}. Alternative frameworks for analyzing privacy have been advanced, most notably privacy as ``contextual integrity'' which focuses privacy analysis on norms of information flows that inhere in various contexts and considers their violation to signal a potential privacy transgression~\cite{nis04}.
Privacy understood as ``a personal notion shaped by culturally determined expectations and perceptions about one's environment'' has been found to require feedback (``informing people when and what information about them is being captured and to whom the information is being made available'') and control (``empowering people to stipulate what information they project and who can get hold of it'') by subjects~\cite{bel93}, mechanisms for ongoing boundary control~\cite{alt77,pal03} and a reciprocity of knowledge and exposure between users and observers~\cite{bel93}. Researchers have explored various mechanisms for representing boundaries~\cite{pal03}, information flows and context to users to enable decisions about the disclosure of personal information.

Related work on location privacy standards in the IETF Geopriv working group takes a play out of the content management environment and builds a mechanism through which users can attach policies to their location information. This turns the standard processes of privacy law and practice that rely on entities providing notice to users through privacy policies on web sites to which users can choose whether to consent on their head. Core elements of the Geopriv approach are  ``limiting the distribution of location information to authorized entities'' and specifying policies for retention. Geopriv also incorporates a feature to support the data protection concern of minimization by providing alternative means for representing location at various levels of granularity with presumptively corresponding levels of privacy and safety concerns. Specifically, Geopriv supports ``geodetic location data (latitude/longitude/altitude/etc.) as well as civic location data (street/city/state/etc.)''~\cite{bar09c}.

The W3C Platform for Privacy Preferences provides a set vocabulary that web sites can use to state their privacy policies in XML format. User agents can read and interpret these policies and present their conclusions to users or, through user delegation, act upon them. P3P increases the transparency of web site practices and facilitates notice and consent by reducing transaction costs associated with privacy communications through standardization and machine readability. The P3P specification includes a vocabulary that constrains the statements that can be made about privacy policies. Conformant P3P polices must provide contact information for the legal entity making the representation of privacy practices, as well as a URI for the human readable policy. They must enumerate the types of data or data elements collected, explain how the data will be used, indicate whether and what access rights are available, and identify the data recipients. The vocabulary also supports statements about the consequences of information use under the policy.

The privacy frameworks and specifications identify important common elements that must be incorporated into a framework to evaluate location information privacy. Specifically, the focus on facilitating end user control over decisions about whether and in what circumstances to disclose location information is facilitated alternatively by user communication of privacy rules and control over location disclosure, or by the principles of notice and consent prior to the disclosure of location information. Consent is made meaningful by required disclosures about the purpose of the request, intended secondary uses or disclosures and retention periods. Knowledge, consent and feedback is further supported by mechanisms that allow users access to the data maintained about them and the ability to amend or delete it. Research suggests that feedback improves user comfort and allays privacy concerns~\cite{tsa09}.

Based upon a review of existing frameworks and prior standards that address privacy we suggest the following framework of common questions for analyzing the privacy support provided by a technical standard. We frame the analysis in terms of \emph{support} to acknowledge both the range of influence that technical standards can have on privacy as well as the power of other constraints, such as law. In considering the impact of technical standards we in no way mean to suggest that the standard alone can or ought to independently determine location privacy. In fact, the determination of where the regulatory hand-off between technological, legal, market and normative constraints occurs is of utmost importance. In an area with complex and context-dependent laws and norms about information flows, standards will do best to enable variation rather than dictate a single outcome~\cite{cla02}.

\begin{itemize}
\item \emph{Appropriateness:} Is the collection of location information appropriate given the context of the service or application?
\item \emph{Minimization:}  Is the minimum necessary granularity of location information distributed or collected?
\item \emph{User Control:} How much ongoing control does the user have over location information? Is the user a passive receiver of notices or an active transmitter of policies?  Are there defaults?  Do they privilege privacy or information flow?
\item \emph{Notice:} Can requesters transmit information about their identity and practices?  What information is required to be provided to the user by the requesting entity? What rules can individuals establish, attach to their location information and transmit?  Is there a standard language for such rules?
\item \emph{Consent:} Is the user in control of decisions to disclose location information? Is control provided on a per use, per recipient or some other basis? Is it operationalized as an opt-in, opt-out or opt model?
\item \emph{Secondary Use:} Is user consent required for secondary use (a use beyond the one for which the information was supplied by the user)? Do mechanisms facilitate setting of limits or asking permission for secondary uses?
\item \emph{Distribution:} Is distribution of location information limited to the entity with whom the individual believes they are interacting or is information re-transmitted to others?
\item \emph{Retention:} Are timestamps for limiting retention attached to location information? How can policy statements about retention be made?
\item \emph{Transparency and Feedback:} Are flows of information transparent to the individual?  Does the specification facilitate individual access and related rights?  Are there mechanisms to log location information requests and is it easy for individuals to access such logs?
\item \emph{Aggregation:} Does the standard facilitate aggregation of location information on specific users or users generally?  Does the specification create persistent unique identifiers?
\end{itemize}

Collectively, answers to these questions provide insight into the capacity of the specification to support interactions under a range of privacy conditions.  In the following sections we consider how these questions are addressed by the W3C Geolocation API specification and its current implementations.

\section{W3C Geolocation API Specification}\label{api-summary}

The W3C Geolocation API~\cite{geolocationapi} provides a simple, high-level JavaScript API to allow web sites to request location information --- primarily latitude and longitude coordinates --- from web browsers, whether on a mobile phone or a laptop computer or any other Web-capable device. The API itself is agnostic to how the browser or device determines the current location: a phone or other mobile device might use a Global Positioning System (GPS) receiver, while a laptop's location might be triangulated from nearby Wi-Fi networks or inferred from its IP address.

Web sites may either request a ``one-shot'' location --- commonly used to locate the user on a map or show nearby points-of-interest --- or register with the browser to receive regular updates, which may be used to give directions as the user moves through a city, for example.

In each case, precise location information (latitude and longitude, and sometimes also altitude, heading and speed) is provided to the calling JavaScript code along with accuracy information (measured in meters and corresponding to a 95\% confidence level). This JavaScript code runs inside the browser on the user's own machine, but in most cases it immediately communicates the user's location to the hosting web server or some third-party server (like Google Maps, for example) using AJAX or an equivalent method.

The specification provides \emph{normative} requirements for both user agents (the web browser) and recipients (the requesting web site) in order to protect user privacy. (Normative requirements are binding for any implementation that conforms with the standard.)
Requirements may be functional --- governing the actual API function calls between the browser and web site --- or non-functional --- obligations for the browser or web site that don't directly affect the operation of the API.
The specification also includes non-normative (that is, non-binding) recommendations and leaves other questions completely open to the implementer's discretion.
Both the choice of normative requirements (or the lack thereof) and whether those requirements are functional or non-functional affect user privacy in practice.

\subsection{User Agent Requirements}\label{api-useragents}

Most of the normative privacy requirements on user agents (typically web browsers) are functional, specifying down to the ordering of distinct steps the algorithm for returning the user's location.
Most notably, the user agent must, in most cases, get the user's express permission to send the device's location to a particular web site before initiating a process to obtain a cached or new location.
This directly addresses the \emph{consent} question of our privacy framework: users are given yes-or-no control before their information is obtained or revealed.
The specification requires that the user agent identify the web site by its \emph{document origin}, as defined by HTML5~\cite{html5}.
The document origin specifies which web site is theoretically responsible for requesting the location, although the JavaScript code itself may have come from a third-party server (web sites commonly include JavaScript from Google in order to do basic traffic monitoring, for example) and the JavaScript code may communicate the information to a different, third-party server without the user's knowledge or consent.
Providing the document origin gives the user some \emph{notice} of which entity is requesting their information (though it is silent with respect to their practices), but doesn't address questions of \emph{distribution}, since users have no assurance about which entities may ultimately receive their location.

In certain cases, the user agent does not have to obtain express permission from the user before obtaining and sending location. First, if the user agent has a ``prearranged trust relationship'' with the requester --- the example given in the spec is the case of a VoIP telephone with an E911 function to inform emergency services of the user's location --- then the user agent need not show a user interface or request permission. No implementing web browser (Firefox, Mobile Safari, Opera) currently has default ``prearranged trust relationships'' with law enforcement or any other web site, and it is not clear how this phrase, not explicitly defined in the spec, will be interpreted by user agents.
The privacy impacts of such pre-arranged relationships would turn on their \emph{appropriateness}: users are likely to expect and appreciate automatic transmission of their location when requesting emergency services but would consider it inappropriate for their user agent to automatically track their location history while browsing in order to provide targeted advertising.

Second, users may persist their permissions for a particular web site to access their location. Commonly when a user is asked to give permission to a web site, the browser also presents an option to remember their answer (whether negative or affirmative). If a user asks the browser to remember that they have given permission to a web site, then the browser is not required to present any user interface or obtain permission the next time the user accesses that web site and the web site requests the user's location. These permissions persist even if the purpose for which the web site is collecting location information or its practices of secondary uses, retention and re-distribution change; in fact, even a conscientious requesting web site has no way of signaling to the user agent that their policies have changed and the user should be prompted again.

The specification provides no requirements on whether persistent permissions should expire over time and browser implementations vary on this point. User agents are required to let users manually revoke persisted permissions and are advised ``to provide easy access to interfaces that enable revocation''~\cite[Section~4.3]{geolocationapi}, but browser implementations vary on how easy this is in practice. For more details, see Section~\ref{clients}.

Web sites may either request an old position that the user agent has cached, or insist that a new location be obtained. The specification provides no requirements or recommendations on how long browsers may or should cache a user's location, and provides no method for users to see what cached location (which may not be their current location) will be provided to the requesting code.
Nor can users override or control whether new or cached locations are provided.  Enabling users to inspect location data before sending it could support the \emph{transparency} of the system and let users censor sensitive locations.

The specification provides a non-normative (that is, non-binding) section of ``Additional implementation considerations'' for user agents which recommends (though does not require) that web browsers ``enable user awareness of location sharing''~\cite[Section~4.3]{geolocationapi}. Such features would also support \emph{transparency}, but, as described in Section~\ref{clients}, web browsers don't currently provide any ambient feedback to the user to remind them when their location is being shared.

Finally, the current draft of the API provides no requirement for the ability of clients to send a ``fuzzed'' location or anything less granular than the exact latitude and longitude (just the city and state, for example).
Yahoo's Fire Eagle location broker provides this functionality at eight different levels of granularity\footnote{\uri{http://fireeagle.yahoo.net/developer/documentation/location}} and the Firefox Geode feature (a forerunner to the W3C API) provided four different levels of granularity.\footnote{\uri{http://mozillalabs.com/blog/2008/10/introducing-geode/}}
The Working Group is considering something like this functionality for the second version of the API,\footnote{\uri{http://www.w3.org/2008/geolocation/track/products/2}} but this has been ruled out of scope for the current version over concerns that naive fuzzing algorithms could be circumvented by making multiple requests over time.\footnote{\uri{http://www.w3.org/2008/geolocation/track/issues/18}}
As a result, the API cannot fulfill the privacy principle of \emph{data minimization}: even in cases where the web site may not need the exact location, the API always requires that the precise latitude and longitude be sent. The API provides users a choice between allowing real-time tracking of precise location or being unable to use location-based services.

\subsection{Requesting Web Site Requirements}\label{api-websites}

Although the primary audience for (and the primary authors of) the specification are the various browser makers (Apple, Google, Mozilla and Opera will have to implement the API as part of their browsers), the specification also puts normative requirements on requesting web sites that use the API in their JavaScript code~\cite[Section~4.2]{geolocationapi}.
Web sites that do not satisfy the ``Privacy considerations for recipients of location information'' are not considered conformant with the specification,\footnote{\uri{http://lists.w3.org/Archives/Public/public-geolocation/2009Oct/0009.html}} but as a practical matter the W3C lacks any power to enforce web site practices, and thus there is no clear negative consequence for non-conformance.

Nevertheless, the W3C specification insists that web sites which request visitors' locations using the API meet strict requirements for notice, consent and usage of visitor location information. Without the express permission of the user, web sites must not use location information for purposes beyond which it was provided to them, retain location information longer than permitted or retransmit the visitor's location. Also, web sites must ``clearly and conspicuously disclose'' that they are collecting visitor location data, how long it is kept, how it is secured, whether it is shared, and how visitors may access, update or delete it.
These requirements directly and explicitly address the \emph{appropriateness}, \emph{notice}, \emph{consent}, \emph{secondary use}, \emph{distribution}, \emph{retention} and \emph{transparency} aspects of user privacy.

But though these requirements are \emph{normative} sections of the specification, they are not \emph{functional} requirements that directly influence how the API works.
None of these notices are communicated as part of API calls, and none of these requirements are enforced by the user agent (as a practical matter, it is impossible to enforce them, because the API does not provide any way in which this enforcement could be supported).
Instead, web sites are expected to use the HTML content of their own pages to make details about collection, usage, storage and access clear to their visitors. The specification does not detail any particular interface or language requirements and no de-facto standards exist.
Web sites vary in their implementation of these rules and very often fall short; see Section~\ref{services} for an initial survey of sites using the API.

The W3C Geolocation Working Group did consider multiple proposals that would have added functional requirements for requesting web sites: making either transmitted privacy preferences from the user\footnote{\uri{https://datatracker.ietf.org/liaison/486/}} or notification of intended usage from the web site\footnote{\uri{http://lists.w3.org/Archives/Public/public-geolocation/2009Mar/0131.html}} part of the API function calls.
The Geopriv proposal~\cite{rfc3693,rfc3694}, designed and endorsed by the IETF,\footnote{\uri{http://www.ietf.org/dyn/wg/charter/geopriv-charter.html}} would have allowed users to attach a set of personal rules to their location information so that requesting web sites would be informed in a machine-readable way whether the user permitted long-term storage or re-transmission of their location information.
Such a proposal would also have changed \emph{user control} of location information from passive (users receiving and accepting notices) to active (users transmitting their own preferences).
In a contentious vote, the Working Group chose not to adopt the Geopriv proposal for the current draft of the specification, with opponents arguing that it was too complex to be widely used or understood by web site developers or novice users.\footnote{\uri{http://www.w3.org/2008/12/08-geolocation-minutes.html\#item02}}
The Working Group leadership also rejected proposals to have websites send an intended usage notification and fields specifying retention time and whether or not information would be retransmitted, which would have standardized and enforced notice of \emph{secondary use}, \emph{distribution} and \emph{retention} policies.
Opponents feared that such a system would erode user trust in web browsers since web site privacy practices could not be enforced by the browser itself,\footnote{\uri{http://www.w3.org/2008/geolocation/track/issues/10}} while proponents (including one of the authors) suggested that this need not be any different than displaying other third-party information.\footnote{\uri{http://lists.w3.org/Archives/Public/public-geolocation/2009Apr/0054.html}}

\section{Requesting Web Site Implementations}\label{services}

Although the W3C Geolocation API is only in draft form, its presence in popular browsers (iPhone's Mobile Safari and Firefox 3.5+) has led some web sites to make use of it already. We sought to catalog and evaluate the popular web sites that use the API today and their privacy policies: both whether they satisfied the ``Security and privacy considerations'' section of the API and whether they made their practices clear to the user up front, before invoking the API. The normative, non-functional requirements put on web sites by the API specification employ a traditional model of notice and consent and rely on web sites to proactively provide clear, conspicuous and comprehensive notice; we sought to investigate whether such notice was actually present.

\subsection{Method}

To obtain a corpus of web sites using the API, we used the 80legs\footnote{\uri{http://80legs.com}} distributed web-spidering service to crawl 11 million URIs and search their JavaScript code (both inline in HTML code and in separate JavaScript files) for references to the API. 80legs does not choose URIs randomly from the trillions of URIs on the Web today; we seeded our search with the 10,000 domains that receive the most traffic according to \code{alexa.com} and our 80legs code followed links on those pages to other web sites.

On obtaining results from the crawl, we manually investigated each web site that matched our search to determine whether it was actually using the API, what it apparently used the API for, whether the user was informed of the site's practices with user location information before being prompted for permission, whether user location information was mentioned in the site's Privacy Policy and whether the relevant Privacy Policy was easily accessible (with a single click, for example) from the page where the user was prompted. We also investigated one aspect of \emph{user control} and \emph{distributional} norms: whether users could inspect their location information and then choose whether to submit it to the requesting site.

We removed from the list any page that was clearly just a test page (like our own) to experiment with the API, any page that was not actually using the API (some pages were simply describing the API without using it themselves and others had the code necessary for the API but did not activate it) or any page we couldn't test (some were behind membership walls). We added some sites to our list that were not found in the automated crawl that were reported to us by participants in the Geolocation Working Group.

The full list of sites is available online at \uri{http://npdoty.name/location/services}. Since this was compiled during the Fall, undoubtedly some sites that we crawled have since added support for the API; our analysis of the privacy practices of this list is current as of February 2010.

\begin{figure}[!thb]
\begin{center}
\includegraphics[width=0.7\columnwidth]{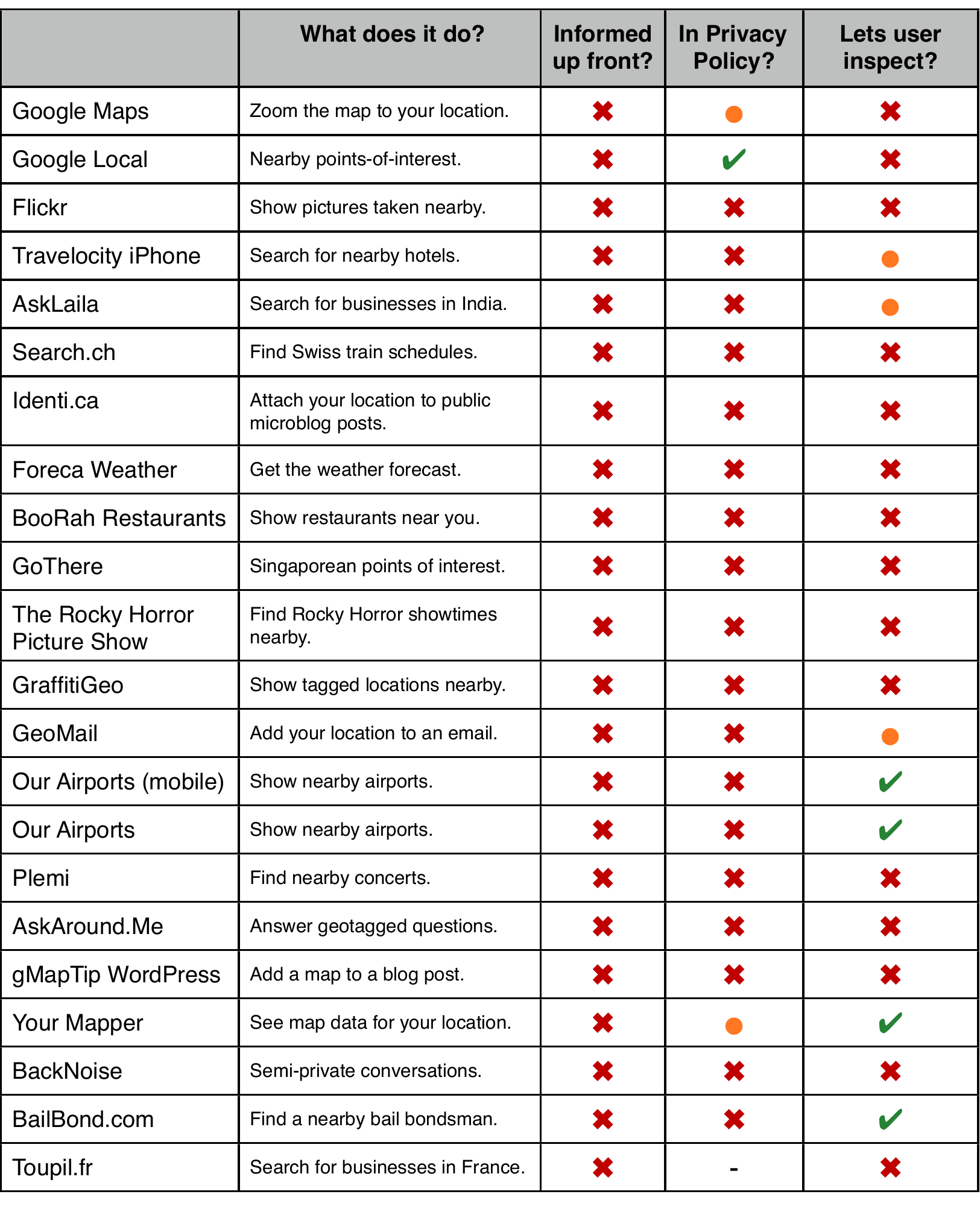}
\caption{\emph{Web sites using the W3C Geolocation API.} For a complete and up-to-date list, see \uri{http://npdoty.name/location/services}}\label{table-location-sites}
\end{center}
\end{figure}

For several reasons, our list should not be considered a comprehensive or exhaustive search of web sites using the API. 11 million URIs is a tiny fraction of the Web and our crawling algorithm was not exhaustive of all pages on each of the high-traffic domains it was seeded with. Some web sites only transmit code that uses the API if the browser is identified as running on a mobile phone: the 80legs crawler is not. Furthermore, some web sites use the Robot Exclusion Standard to specifically prevent automated crawlers from accessing JavaScript files. Finally, it is possible that some code that our crawler accessed might not have used the particular string we searched for (``navigator.geolocation'') in calling the API. But though our list of 22 implementing web sites is not exhaustive, we do believe that it provides a useful sample of API usage that users are likely to encounter.\footnote{We're currently exploring techniques that would allow us to make an even broader and more comprehensive search of web sites; if successful, we will update our list and analysis with the additional data.}

\subsection{Results}

Analysis of the 22 implementing web sites revealed a variety of practices in notice to the user and usage of location data. But for the most part, current implementing web sites do not ``clearly and conspicuously'' disclose the purpose and practices of their collection of users' location information.

Sites are listed in Figure~\ref{table-location-sites} ordered by their Alexa ranking; the bottom seven sites in the list are unranked.  No site informed users of their information practices before requesting the user's location (the first column).  In the second column, sites where the privacy policy did not explicitly cover location information are noted with a red X, while sites with a clear privacy policy easily accessible from the requesting page are noted with a green checkmark.  In the third column, sites that allow inspection of location information before submitting are noted with a green check and sites that wait to submit location information until the user submits a form are noted with an orange circle.

Out of 22 instances, not a single web site informed users of their privacy practices with respect to collected location data up front, that is, before they were presented with a prompt for their location. As a result, we suspect that virtually no users encountering the W3C Geolocation API are fully informed about the requesting site's information practices when they decide whether or not to reveal their location. Nine sites (41\%) presented the prompt immediately on loading a page, without a user even pressing a button to initiate the action.

Furthermore, only four of the 22 sites (18\%) explicitly mentioned the collected location data in their privacy policy (if a privacy policy even existed), making it difficult for even a determined user to find out policies on \emph{distribution} or \emph{retention}. Only one site (the mobile version of Google Local) had a privacy policy with a clear description of the collected location data available within a single click.

In the case of the microblogging site Identi.ca, logged-in users are immediately presented with the location prompt on loading every page with no explanation of what their location will be used for, casting doubt on the \emph{appropriateness} of the request. (The prompt even appears on their privacy page, which does not mention location information.)  Users who do provide their location have it silently and automatically attached to the posts they make on the site, which are published (alongside their usernames) publicly for everyone to see, a form of \emph{distribution} they may not expect.

There was one encouraging sign for privacy protection: seven of the sites (32\%) did not immediately submit the user's location back to the originating web server once it was received. Instead, they entered the received location data into a form field on the page itself; location information would only be sent to the server if and when the user submitted the form by clicking a button (with labels like ``Search''). In some of these cases (four of the seven), the user can even inspect or modify the location data before it is sent: to correct a mistake that their browser made, censor a sensitive location that they do not want to share, decrease the precision of the location data being sent or simply to be aware of what personal information is being provided to the web site. For example, pilots who use OurAirports.com can click ``Use my current location" to bring up the Geolocation API prompt, after which their latitude and longitude is presented in plain text in the search box, where a pilot can edit it before finally clicking ``Go!'' to find nearby airports. Such techniques, though uncommon in our survey, match common expectations for how forms on the Web work, enable real \emph{transparency} and maximize \emph{user control}.

Of course, because the API currently returns location data exclusively in latitude and longitude coordinates, very few users (other than pilots) can make sense of coordinate data they are about to send, even if they are provided an opportunity to inspect or change it. BailBond.com and YourMapper.com take the helpful step of \emph{reverse geocoding} the user's location, translating it into a human-readable city and state, but this requires first transmitting the user's coordinates to a service (GeoNames and Google, respectively) to do the geocoding without the user's notice or consent.

\section{Client Implementations and Location Providers}\label{clients}

We have not completed a systematic analysis of supporting Web browsers, but popular implementations (Firefox 3.5+ and Mobile Safari 3.0+)\footnote{Beta and experimental versions of the Opera browser also support the API, but this functionality hasn't been released as part of the main product. Google Chrome and other browsers with the Google Gears plugin support similar functionality but with a distinct API.} do vary. All browsers explicitly request the user's permission through a UI before revealing location. But persistence of permissions (after which the user is not prompted anymore) varies: in Firefox, users check a box labeled ``Remember for this site''; on the iPhone, the user's preference is remembered after two consecutive answers.

The specification requires that user agents provide a way to revoke these persisted permissions and advises that they ``provide easy access to interfaces that enable revocation''. In Firefox, this requires returning to the page in question (risking that the site will again automatically obtain your location) and viewing the permissions for the page in ``Page Info''. On the iPhone, persisted permissions are automatically reset after 24 hours. Users can manually reset location warnings for all applications on the phone through a series of menus in the phone's settings, but this resets all location information permissions granted by the user; there is no way to revoke location permissions for a single application or site. Once the user has given permission for Mobile Safari (which in this logic is just a single application providing access to location information) to access their location, the iPhone remembers any previously set site-specific permissions.

Although web browsers ``are advised to enable user awareness of location sharing'' neither Firefox nor Mobile Safari provide any sort of ambient notification to the user when location data is automatically sent to a web site for which permissions are persisted.  Similar functionality does exist today at the operating system level: the Windows 7 Location Platform\footnote{\uri{http://www.microsoft.com/whdc/device/sensors/}} provides unobtrusive taskbar notifications when an application accesses the user's location and also lets users view a retrospective log of location activity (which applications accessed the user's location and when).\footnote{\uri{http://msdn.microsoft.com/en-us/library/dd756632(VS.85).aspx}}

Finally, although the specification is agnostic to how implementing user agents determine the user's location, the details can affect user privacy. For example, Firefox sends a list of the user's Wi-Fi SSIDs and signal strengths to Google Location Services in order to estimate location, and Google stores that list along with the user's IP address and a unique client identifier.\footnote{\scriptsize\uri{http://www.google.com/privacy-lsf.html}} On the iPhone (and in the Opera web browser), Wi-Fi triangulation is done through Skyhook Wireless, which has a similar privacy policy.\footnote{\scriptsize\uri{http://www.skyhookwireless.com/howitworks/privacypolicy.php}}  There are several different techniques that could be used for geolocating a particular Web-based device:

\begin{itemize}

\item\emph{Geolocation by IP Address:} If the IP address of a device is known, the general geographic area can be inferred. The result varies widely in accuracy and is based on the assumption that an IP address belongs to an IP address block that can be associated with a physical location. This assumption can be problematic in a variety of settings, such as with dynamic IP addresses, VPNs, proxy servers, NAT, and other network-level technologies that influence IP address assignment.

\item\emph{Geolocation from Wi-Fi Networks:} For devices that have Wi-Fi access, the SSID (a network identification code) of nearby networks can be used to lookup the geolocation of that wireless network in a database of SSIDs. This method is often reliable, but suffers from the fact that SSIDs are not unique and can change position.

\item\emph{Geolocation by Cell Tower Triangulation:} For devices using mobile phone networks, cell tower IDs can be used to determine location by triangulation (using various IDs and the signal strength to calculate the most likely location). The accuracy of this method varies with the cell tower density.

\item\emph{GPS:} An increasing number of devices have GPS receivers built-in, which produce very accurate location measurements. GPS has the disadvantages of being less reliable indoors, being affected by the visible sky in a given location and potentially taking a long time to fix on a location.

\item\emph{Manual Input:} Users can also manually specify their location, either selecting from a predefined list, or using interfaces with map views which might be initialized by a different location technology. (No browser implementations of the W3C Geolocation API support manual input today.)

\end{itemize}

Because the specification is agnostic to how location is provided, user agent implementations could allow users to choose between different providers or different methods under different circumstances, to obtain more accurate location information in certain circumstances or to hide sensitive locations from a particular provider.
For example, Firefox could provide an interface that let users manually specify their location (Fire Eagle provides such functionality for its location broker service) or the iPhone could provide a setting to only use GPS for location services (which would decrease their speed and accuracy in some circumstances, but prevent Skyhook or Apple from being informed of the user's location).
But as a matter of fact, Google and Skyhook act as the provider for the vast majority of users of the W3C Geolocation API today.
Using large, centralized location providers not only is more economical for device and browser manufacturers then implementing such a service in-house, it also allows providers to use the large volume of location requests to continuously improve and correct their databases, for example, to detect when a particular Wi-Fi access point has suddenly moved.
But such centralization has the side effect of enabling large-scale \emph{aggregation} of user location data: although a user might reveal his location to several different location-based services throughout the course of a day, the user reveals his location to the same location provider each time, which can produce a full, sensitive and personally-identifying location history for a pseudonymous user.
The privacy risks of large-scale aggregation on the Web are not new: the KnowPrivacy project~\cite{gom09} and a study by Krishnamurthy and Wills~\cite{kri09} illustrate some of the possible threats. Recent regulator and public concern about the retention of search queries is also applicable to these backend location providers; although they collect different data, aggregated location data raises not only a similar privacy concern but also a safety concern, since a large number of individuals may be physically located with such a dataset.

\section{Try It Yourself}\label{try-it-yourself}

Since a handful of browsers and web sites are already using the draft W3C Geolocation API, it is not difficult to try this new functionality for yourself. Use an iPhone (with v3.0 or higher software) or a recent version (3.5 or higher) of the free Firefox browser and navigate to our test page at \uri{http://npdoty.name/location}.

On loading the test page you can immediately see the accuracy of the legacy, non-JavaScript method for determining the user's location. Just by receiving the request, a web site can guess location (sometimes to the city level of precision) from the user's IP address. By clicking ``Find my location'' you can see the permission dialog that your browser uses to obtain your express consent and, on giving your permission, you can see all the information that your browser returns about your location. The amount and accuracy of data will depend on your browser, how it obtains your location and where you are. At different times, your browser might return location data with accuracy to 20,000 meters or to 20 meters.

\section{Future Developments}\label{future}

The W3C Geolocation API is only one example of how potentially sensitive personal information is made available on the Web. The current trend towards increasingly capable mobile devices and the mobile Web suggests that these privacy issues will become increasingly common and important in the near future. For example, the BONDI\footnote{\uri{http://bondi.omtp.org/}} initiative standardizes a set of access technologies for mobile devices which go well beyond just location, including access to address books and calendar information. Mobile devices typically contain substantial private information, and the small form factor makes it challenging to design and implement complex access control and management interfaces.

In the W3C, the \emph{Device APIs and Policy Working Group} (DAP)\footnote{\uri{http://www.w3.org/2009/dap/}} is tasked with addressing these challenges. The group's mission is ``to create client-side APIs that enable the development of Web Applications and Web Widgets that interact with devices services such as Calendar, Contacts, Camera, etc.'' and to ``produce a framework for the expression of security policies that govern access to security-critical APIs.'' The working group has not yet produced any public documents, but members have made it clear that they're closely following the discussion of privacy in the Geolocation API as they consider a broader set of APIs.\footnote{\uri{http://lists.w3.org/Archives/Public/public-geolocation/2010Jan/0002.html}}


The details of the W3C Geolocation API and its privacy support may also affect law and regulation. The \emph{Federal Trade Commission (FTC)} has expressed interest in the potential harms to consumer privacy that can arise from mobile devices and location-based services as part of its recent ``Exploring Privacy'' roundtable series.\footnote{\uri{http://www.ftc.gov/bcp/workshops/privacyroundtables/}} Meanwhile, the \emph{House Committee on Energy and Commerce} recently held a hearing on ``The Collection and Use of Location Information for Commercial Purposes'' which may inform broad federal privacy legislation. Both the FTC and Congress are likely to take current practices and any industry self-regulation into account when determining whether additional regulation or legislation is necessary.

Figuring out where is best to address privacy concerns --- in specifications, implementations, interfaces, corporate policies or laws --- will be of the utmost importance.  User adoption of location-based services may be slowed if privacy issues are not appropriately considered.

\section{Recommendations}\label{conclusions}

In conclusion, we propose four modifications to the W3C Geolocation API to enable support of the different principles in our privacy framework that appear to be lacking in current implementations.

\begin{enumerate}

\item Any functionality that would enable \emph{minimization} is notably absent from the specification and as a result users must always risk transmitting their precise latitude and longitude location. This increases user risk (of physical harm and of revealing sensitive or personally-identifying information) while in many cases not providing any useful advantage. The specification could either provide a standard for different levels of granularity (perhaps a subset of RFC4119's civic location format) which user agents could expose as a choice to users or add a parameter for ``fuzzing'' latitude and longitude coordinates by a user-specified amount.

\item The specification wisely uses a model of user \emph{consent} and \emph{control} for revealing location information --- location-based services employ a wide variety of methods for how they request, store and distribute user data and the specification should not inhibit this variety. But the model relies on web sites proactively informing users of how they will use, retain and distribute user data and our preliminary exploration shows that web sites consistently fail to do this. Rather than relying on a potentially cumbersome legislative solution to enforce notice, functional requirements could be added to the current specification to allow machine- and human-readable notices to be sent along with each request for user location. This would require web sites to explicitly communicate (at the time of request) to users about \emph{use}, \emph{distribution} and \emph{retention} and, we believe, would increase user control by creating informed consent. \emph{Appropriateness} will always have to be judged on a case-by-case basis, but by providing users with a human-readable explanation of how their information will be used at least provides users with the necessary context to make that decision for themselves.

Alternatively, the specification could (as in the Geopriv specification) support two-way communication about privacy rules.  Allowing users to attach privacy rules to their own location information may lessen the burden on notice and consent processes which can be particularly burdensome given the form factor of mobile devices and the time-sensitive nature of many location-based queries.

\item In many cases the distribution flows of information through the Geolocation API are opaque to the user: users are rarely aware of what exactly their location data consists of, where it's re-transmitted or even, in the case of persisted permissions, if and when it's sent at all.  User preferences may change over time even when the information practices of a particular service are completely constant.  By better supporting \emph{transparency and feedback}, the API could respond to this contextual aspect of privacy.

Since current user agent implementations have chosen not to follow the non-binding advice to enable awareness of location sharing, the specification could require that web browsers, where possible, provide an unobtrusive ambient notice.  Further, the Working Group could learn from the handful of requesting web sites that today let the user inspect their own location information (with or without reverse geocoding) and recommend that web browsers provide that same functionality and control for all the sites their users visit.

\item Although the specification is (and should remain) agnostic to the user agent's particular location provider, it should not ignore the significant risks of \emph{aggregation} among a small number of location providers (currently, Skyhook Wireless and Google). User agents should be encouraged to provide choice between different location providers and required to provide information to users on who their location provider is and what its privacy policy is.\footnote{Firefox already does an excellent job of the latter, explaining in clear terms how the API works and what information is transmitted to Google. See \uri{http://www.mozilla.com/en-US/firefox/geolocation/}.}

\end{enumerate}

\bibliographystyle{dret}
\bibliography{dret,nick}

\end{document}